\documentclass[runningheads]{llncs}
\usepackage{graphicx}
\usepackage{amsmath, amssymb}
\usepackage{hyperref}
\usepackage[utf8]{inputenc}
\usepackage{siunitx}
\usepackage{xcolor}

\begin{document}
	\title{Aligning Subjective Ratings in Clinical Decision Making}

	\author{Annika Pick\inst{1,3,4} \and
		Sebastian Ginzel \inst{1,3,4} \and
		Stefan Rüping \inst{1} \and 
		Jil Sander \inst{1,3} \and
		Ann Christina Foldenauer\inst{2,3} \and
		Michaela Köhm\inst{2,3}}
	\authorrunning{A. Pick et al.}
	%
	\institute{
		Fraunhofer IAIS, Sankt Augustin, Germany \and 
		Fraunhofer IME-TMP, Frankfurt, Germany 
		\email{firstname.lastname@\{iais,ime\}.fraunhofer.de} \and
		Fraunhofer Cluster of Excellence Immune-Mediated Diseases CIMD, Frankfurt, Germany \and
		Fraunhofer Center for Machine Learning, Sankt Augustin, Germany
	}

	\maketitle

	\begin{abstract} 
		
		In addition to objective indicators (e.g. laboratory values), clinical data often contain subjective evaluations by experts (e.g. disease severity assessments).
		While objective indicators are more transparent and robust, the subjective evaluation contains a wealth of expert knowledge and intuition.
		In this work, we demonstrate the potential of pairwise ranking methods to align the subjective evaluation with objective indicators, creating a new score that combines their advantages and facilitates diagnosis.
		In a case study on patients at risk for developing Psoriatic Arthritis, we illustrate that the resulting score (1) increases classification accuracy when detecting disease presence/absence, (2) is sparse and (3) provides a nuanced assessment of severity for subsequent analysis.
		\keywords{Clinical Data \and Ranking SVM \and Data Integration.}
	\end{abstract}
	\section{Introduction}
		In data obtained from clinical studies it is often challenging to determine the dis\-ease status of a patient.
		There are usually multiple ways for a disease to manifest and the ways of manifestation are detected by different examination methods. 
		While the assessment e.g. of swelling of a specific joint is relatively objective and can be performed by a non-specialist as well, the determination of which symptoms are directly caused by  the disease in question including the generation of a complete symptomatic picture is no trivial task.
		
		In general, an overall assessment is approximated by a numerical disease activity (DA) rating specified by the physician. 
		Although such a rating implicitly contains valuable domain expertise when provided by an experienced specialist, the absence of strong diagnostic criteria makes it highly subjective. 
		For example, the well-established Visual Analog Scale (VAS) is a standard measurement instrument for DA assessment of arthritic diseases, but it has been previously shown that VAS differences of as much as~$\SI{15}{\percent}$ fall within the expected variance~\cite{tubach2012minimum}.
		
		In contrast to the DA rating, the variables describing individual symptoms are more objective; however, it takes domain knowledge to weight them correctly and assess the resulting score. 	
		Our goal is to align the DA rating with symptom variables to combine the advantages of the two scores within a single rating. We implement this by learning to predict relative DA rankings from individual symptoms, using a Ranking SVM~\cite{Herbrich1999c}.
		A general challenge is that ratings of the same disease activity may vary widely although the symptoms remain similar. Only patients with significantly larger disease activity express more relevant disease signs leading to substantially different ratings. We address this issue in our method in order to reduce noise.

		To evaluate the resulting model, we test the correlation of the new score to the original DA rating and also its ability to predict the most reliable binary examination result (presence or absence of the disease); if our new score captures the severity of disease more accurately than the raw DA rating, it should also distinguish better between absence and presence of disease. Furthermore, we expect a meaningful model to be sparse.

	\section{Method}\label{method}
		We denote the normalized variables of the dataset describing the clinical features ($m$~symptoms of $n$ patients) as~
		$\mathbf{X}=\{\mathbf{x}_1,\dots,\mathbf{x}_n\}, \mathbf{x}_i \in \mathbb{R}^{m}$ 
		and the label (phy\-si\-cians' ratings of DA for all patients) as~$\mathbf{y} \in \mathbb{R}^n$. 
		We extend the concept of the Ranking SVM~\cite{Herbrich1999c}, which learns pairwise rankings of data points based on pairwise differences.
		In order to account for inaccuracies in $\mathbf{y}$, we adapt the method by training only on pairs where $\mathbf{y}$ differs by at least $\delta \in  \mathbb{R}$:
		\begin{align*}
		\left. \begin{array}{l}
		\mathbf{x}^\text{paired}_{p} = \mathbf{x}_i - \mathbf{x}_j , \\
		\mathbf{y}^\text{paired}_{p} = \operatorname{sign} (\mathbf{y}_i - \mathbf{y}_j) 
		\end{array}\right\} \text{ for } p \in \{ (i,j) \mid i<j \text{ and } \left| \mathbf{y}_i- \mathbf{y}_j \right| \geq \delta \}.
		\end{align*}

		After training a regular SVM on the set of new data points, we obtain a weighting vector $\mathbf{w}$ and the decision function
		\begin{equation*}
		\text{$i$ has lower DA than $j$} 
		\Leftrightarrow \mathbf{w}^\top (\mathbf{x}_i - \mathbf{x}_j) < 0 
		\Leftrightarrow \mathbf{w}^\top \mathbf{x}_i < \mathbf{w}^\top \mathbf{x}_j.
		\end{equation*}
		Thus, for a set of new patients $I$ and their respective symptoms $\mathbf{x}_i, i \in I$, we can calculate $\mathbf{w}^\top \mathbf{x}_i$ for each of them and use this as a new score that maintains an approximate order according to disease activity, as illustrated by the equation above. The approximation is that only pairs of patients with sufficiently different DA are used for training the weights $\mathbf{w}$ of the SVM.
		
	\section{Related Work}
		The original Ranking SVM~\cite{Herbrich1999c} was developed to solve ordinal regression, i.\,e., regression on fixed categories that have an order. 
		There, data points are paired and an SVM learns to arrange them according to the order of their categories. In comparison, our method can also handle a continuous scale without fixed categories by learning only on clearly distinguishable points.

		Another possibility to reflect different levels of confidence regarding the order of different pairs was introduced by Kotsiantis et al.~\cite{kotsiantis2004cost}, where the pairs are weigh\-ted according to their label difference.
		We may evaluate this approach in fu\-ture work, although level of confidence and label difference are not necessarily linearly dependent in our use case. 
		
		The Ranking SVM has also been used in Image Recognition to predict the age of humans based on pictures~\cite{cao2012human}. This problem exhibits similarities to the challenge of DA ratings, since large age difference can easily be identified, but slight differences are hard to detect.
	
	\section{Evaluation}
		We evaluated our method in the context of the skin condition Psoriasis and the risk of developing Psoriatic Arthritis (PsA): Although psoriasis is closely related to PsA (as $\SI{30}{\percent}$ of psoriasis patients will develop PsA over time), a common problem of PsA is the lack of a clear correlation between disease duration, phenotype of skin psoriasis and PsA development. Non-expert physicians frequently misdiagnose PsA due to the multitude of different clinical manifestations and symptoms~\cite{ogdie2020patient}. If diagnosis is late, patients can develop irreversible musculoskeletal damage~\cite{Haroon1045}. 
		
		When PsA first emerged, disease activity scores were often derived from those developed for Rheumatoid Arthritis (RA), e.g. DAS28~\cite{van1990judging}. However, it has become evident today that despite some similarities, the development and the type of manifestation of RA and PsA are very different and therefore they cannot be evaluated by the same scoring systems. This motivates the development of more accurate assessment scores or tools for PsA.
		
		In a prospective study~\cite{koehm2015sat0574}, 391 eligible patients diagnosed with psoriasis vulgaris and the risk for development of PsA were included. 	About~$\SI{35}{\percent}$ of them were diagnosed with PsA during the examinations as part of the study. 

		The presence of PsA is indicated by a binary examination result (PsA detected in physical examination), a rating of disease activity (DA) of PsA on a Visual Analog Scale by the physician, and by various symptom assessments (swollen/tender joints, lab values, etc.). 
		Our goal is to align the subjective disease activity rating with the more objective symptom assessments. 
		Therefore, we measure the correlation between the new score and the original disease activity rating. Additionally, we test the ROC-AUC when using the new score to predict if the disease is present or not, as indicated by the binary examination result.
		
		\begin{figure}
			\centering
			\includegraphics[width=\textwidth]{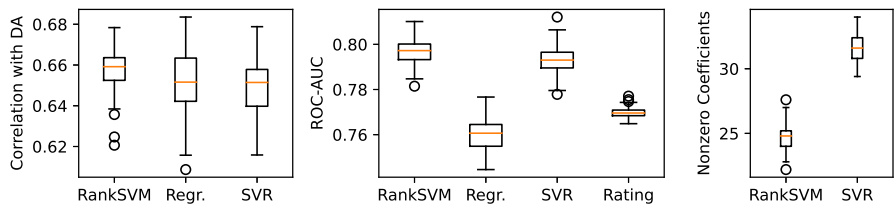}
			\caption{Left: Similar correlation of all models with DA. Center: Scores from Ranking SVM and SVR can predict the binary label better than raw DA ratings; Linear Regression performs worse than raw rating. Right: Ranking SVM needs fewer nonzero coefficients than SVR.} \label{testresults}
		\end{figure}
	
		We compare the Ranking SVM to two baseline models for mirroring the DA: simple Linear Regression and Support Vector Regression (SVR). When testing the ability to predict the presence/absence of disease, we compare the Ranking SVM to the same two baselines as well as to the original, raw DA rating provided by the physician.
		All models are $L_1$-regularized (optimized separately) and linear in order to obtain a sparse and interpretable solution, especially because of the medical context.
		We calculate the DA predictions (via Linear Regression and SVR) and the ranking score (via Ranking SVM) for every patient by using 5-fold cross validation (CV). 
		
		In the Ranking SVM, we set $\delta$ to~$15$ (DA ratings range from 0-100), since this is the minimal clinically important improvement in physician global assessment (absolute value) as determined in~\cite{tubach2012minimum}.
		
		Stability tests (plots omitted due to space limitations) for $\delta$ ranging between 10 and 40 imply that the mean correlation of the new score with DA drops with higher $\delta$, but only by $0.07$ in the tested range. The mean ROC-AUC of detecting disease presence is stable (variation $<0.03$) and the number of coefficients drops from $31$ to $6$ when increasing $\delta$ from $10$ to~$40$. Accordingly, we conclude that the choice is not critical and we chose $15$ as value for $\delta$ due to the semantic reasons explained above.
		
		Figure~\ref{testresults} shows the results of $100$ runs with different CV-splits, since the inhomogeneity of the limited data set led to high fluctuations in performance. 
		All models correlate similarly with DA, but the Ranking SVM is the most stable. 
		With regards to detecting disease presence, both Ranking SVM and SVR enhance the raw DA rating. However, the Ranking SVM performs slightly better and needs~$\SI{20}{\percent}$ fewer nonzero coefficients than the SVR, which is highly relevant since obtaining medical information is costly and time-consuming. A pure classifier SVM was trained to predict the binary label as well (plots omitted due to space limitations); although the mean ROC-AUC was 0.04 higher than the ROC-AUC obtained by using the Ranking SVM, the average correlation with DA was lower by 0.17.

	\section{Conclusion and Future Work}
		We have created a score to reflect disease activity which integrates subjective expert knowledge with (more) objective symptom descriptions and which is able to detect the presence of PsA better than the expert rating alone is. 
		It needs fewer than $25$ nonzero coefficients on average -- compare this to the well-established DAS28 score~\cite{van1990judging}, which needs $58$ attributes (assessment of $28$ joints for swelling and tenderness plus lab value plus physician's rating of DA).
		
		However, this is a work in progress. 
		In future work, we aim to improve the Ranking SVM in several ways. First, by finding a way to integrate pairs below the distance threshold.
		Second, by improving the evaluation methodology, since the current large number of cross-validations and the limited data set make it impossible to set aside a data set of similar size for fitting the sparsity parameters. 
		
		Third, we aim to extend the method to other use cases.
		One of them is that physicians and the patients themselves often rate disease activity differently. For example, Lebwohl et al.~\cite{lebwohl2014patient} show that for Psoriasis patients itching is the factor contributing most to high DA according to their opinion, whereas dermatologists put the highest emphasis on the size and location of skin lesions. The idea is to see this reflected in the weights of the Ranking SVM. Besides that, the alignment of disease activity ratings with symptoms has implications for other complex dis\-eases with an activity rating as well, e.g. multiple sclerosis or schizophrenia.

	\section{Acknowledgements}
	This publication is a joined work between the Fraunhofer Cluster of Excellence for Immune-Mediated Diseases and the Fraunhofer Center for Machine Learning within the Fraunhofer Cluster for Cognitive Internet Technologies. It has also been partially funded by the Federal Ministry of Education and Research of Germany as part of the competence center for machine learning ML2R (01IS18038B).
	
	We thank the anonymous reviewers for their valuable comments and suggestions.
	%
	%
	%
	
	\nocite{*}
	\bibliographystyle{splncs04}
	\bibliography{pharML}

\end{document}